\documentclass[reprint, letterpaper, english, amssymb, nolongbibliography, noeprint, prb, superscriptaddress]{revtex4-2}  
\usepackage{mathptmx}
\usepackage[T1]{fontenc}
\usepackage[utf8]{inputenc}
\usepackage{babel}
\usepackage{microtype}
\usepackage{hyperref}
\hypersetup{colorlinks=true,linkcolor=blue,citecolor=blue,urlcolor=blue,filecolor=blue}
\usepackage{booktabs}

\usepackage{graphicx}
\usepackage{tabularx}
\usepackage[FIGTOPCAP,nooneline,raggedright,normalsize]{subfigure}
\usepackage{xcolor}
\usepackage{placeins}

\usepackage[babel]{csquotes} %Einstellung von Anführungszeichen, Befehl \enquote{}
\usepackage{amsmath}
\usepackage{bm}
\usepackage[locale=US, separate-uncertainty  = true, per=frac,fraction=nice]{siunitx}
\DeclareSIUnit\au{a{.}u{.}}

\usepackage{mdwlist}
\usepackage{upgreek}
\usepackage{chemformula}	%chemische Summenformeln, Reaktionsgleichungen
\usepackage{chemfig}		%chemische Strukturformeln
\usepackage{nicefrac}

\begin{document}					
\title{Momentum Density Spectroscopy of Pd: Comparison of 2D-ACAR and Compton Scattering Using a New 1D-to-2D Reconstruction Method}

\author{Josef Ketels}
\email{josef.ketels@frm2.tum.de}
\affiliation{Physik-Department, Technische Universität München, James-Franck-Str. 1, 85748 Garching, Germany}
\author{David Billington}
\affiliation{School of Physics and Astronomy, Cardiff University, Queen's Building, The Parade, Cardiff, CF24 3AA, United Kingdom}
\author{Stephen B. Dugdale}
\affiliation{H. H. Wills Physics Laboratory, University of Bristol, Tyndall Avenue, Bristol, BS8 1TL, United Kingdom}
\author{Michael Leitner}
\affiliation{Physik-Department, Technische Universität München, James-Franck-Str. 1, 85748 Garching, Germany}
\affiliation{Heinz Maier-Leibnitz Zentrum (MLZ), Technische Universität München, Lichtenbergstr. 1, 85748 Garching, Germany}
\author{Christoph P. Hugenschmidt}
\email{christoph.hugenschmidt@frm2.tum.de}
\affiliation{Physik-Department, Technische Universität München, James-Franck-Str. 1, 85748 Garching, Germany}
\affiliation{Heinz Maier-Leibnitz Zentrum (MLZ), Technische Universität München, Lichtenbergstr. 1, 85748 Garching, Germany}

\begin{abstract}
Two-dimensional angular correlation of annihilation radiation (2D-ACAR) and Compton scattering are both powerful techniques to investigate the bulk electronic structure of crystalline solids through the momentum density of the electrons. Here we apply both methods to a single crystal of Pd to study the electron momentum density and the occupancy in the first Brillouin zone, and to point out the complementary nature of the two techniques. To retrieve the 2D spectra from 1D Compton profiles, a new direct inversion method (DIM) is implemented and benchmarked against the well-established Cormack’s method. The comparison of experimental spectra with first principles density functional theory calculations of the electron momentum density and the two photon momentum density clearly reveals the importance of positron probing effects on the determination of the electronic structure. While the calculations are in good agreement with the experimental data, our results highlight some significant discrepancies.

\end{abstract}

\maketitle		

\section{Introduction}

Fundamental physical material properties such as magnetism, electrical conductivity or topological effects are determined by the electronic structure. In order to improve the understanding of the various characteristics of materials, the experimental determination of the electronic structure, and particularly the Fermi surface of metals \cite{Dugdale2016:Life}, is of high importance. Angle-resolved photoemission spectroscopy (ARPES) has established itself as one of the most popular and powerful tools in this regard, especially in the investigation of 2D electronic systems as it directly probes the 2D band structure \cite{Damascelli2004:Probing}. In contrast, quantum oscillatory techniques e.\,g. exploiting the de Haas–van Alphen (dHvA) effect, directly probe the bulk Fermi surface but put strong restrictions on the ambient conditions as they require high magnetic fields, very low temperatures and crystals with very low disorder \cite{Onsager1952:Interpretation}.

Compton scattering and the measurement of the angular correlation of electron-positron annihilation radiation (ACAR) are two complementary techniques which directly measure the electron momentum density (EMD) in the bulk \cite{Dugdale2014:Probing}. While Compton scattering delivers the 1D projection of the EMD, 2D-ACAR measures the integral along a single direction, i.\,e. the 2D projection of the two-photon momentum density (TPMD) which is closely connected to the EMD, but includes the influence of the positron. Both have low demands on ambient conditions, such as magnetic fields and temperature, meaning that they can easily access large regions of the phase diagram even across phase transitions. While Compton scattering is equally sensitive to \textit{all} electrons, 2D-ACAR is more sensitive to the valence electrons, which are most relevant for the chemical and physical properties, due to the repulsive Coulomb interaction between the positrons and the atomic cores. Furthermore, both techniques enable the analysis of magnetic materials by spin-resolved measurements. There are numerous examples of the application of ACAR and Compton scattering on different material classes including elemental crystals \cite{Ceeh2016:Local, Weber2017:Electronic, Dugdale1998:Fermiology, Dixon1998:Spin, Brancewicz2013:Compton}, alloys \cite{Wilkinson2001:Fermi, Robarts2020:Extreme}, magnetic compounds \cite{Hanssen1990:Positron, Duffy2000:Induced, Deb2001:Magnetic, Major2004:Direct, Dashora2011:Temperature, Mizusaki2012:Spin, Haynes2012:Positron, Weber2015:Spin,  Billington2015:Magnetic}, heavy fermions \cite{Rusz2005:Nature} and superconductors \cite{Smedskjaer1988:fermi,Mijnarends1991:Positron,Sakurai2011:Imaging}. A review on the applications of both techniques can be found in Ref. \cite{Dugdale2014:Probing}. Although there have been measurements of both 2D-ACAR and Compton scattering on the same material (for example, Mg \cite{Nakashima1992:Study,Brancewicz2013:Compton} or Y \cite{Dugdale1997:Direct,KontrymSznajd2002:Electron}) it is very unusual for measurements to be made on the same physical sample. 

While calculating 1D projections from a 2D distribution is mathematically trivial, the inverse transformation is more complex. Usually, the 1D-to-2D reconstruction problem, e.\,g. in Compton scattering, is solved by methods either inspired by the analytical inversion of the Radon transformation or by series expansion, e.g. the Cormack method \cite{Kontrym-Sznajd2009:Fermiology}.  In this paper, we present a more general approach to solve this inverse problem. It employs linear matrices to model the experiment and a quadratic regularization functional to reduce experimental noise in the reconstruction. Thus, the solution of the reconstruction problem is the solution of a linear system of equations, which can be found by direct inversion. This is why we call our method the Direct Inversion Method (DIM). Similar approaches have been employed for 2D-to-3D reconstruction in 2D-ACAR \cite{Cooper2004:X, Fornalski2010:Application, Pylak2011:Reconstruction, Pylak2011:Electron, Weber2015:Spin, Weber2017:Electronic}. However, as all of those approaches use an entropy-like, non-linear regularization functional, iterative algorithms are required to find a reconstruction. In contrast, our DIM uses first and second derivative regularization (a general case of Thikonov regularization \cite{Tikhonov1963:Solution}) to reconstruct a large area of interest without iterations. This regularization functional was also used in the algorithm proposed by Leitner \textit{et al.} \cite{Leitner2016:Fermi} for the direct reconstruction of the 3D Fermi surface from 2D data. To benchmark the new approach, it is compared with the Cormack method as modified by Kontrym-Sznajd \cite{Kontrym-Sznajd1990:Three}.

\begin{figure}

\subfigure[]{\includegraphics[width=0.8 \columnwidth]{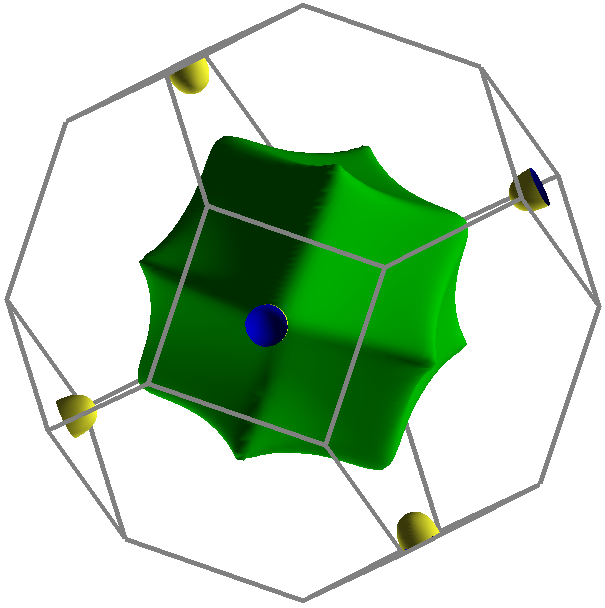}}
\subfigure[]{\includegraphics[width=0.7 \columnwidth]{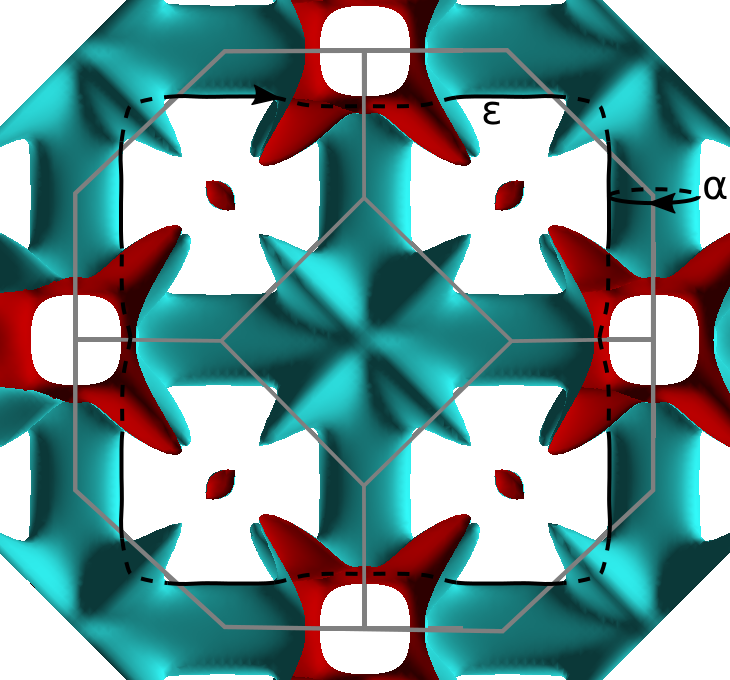}}
\caption{(a) $\Gamma$-centered electron sheet and $X$-hole pockets of the Fermi surface of Pd in the first Brillouin zone (b) Open-hole sheet and $L$-hole pockets of the Fermi surface. Additionally, the dHvA $\alpha$ and $\epsilon$-orbit is shown. The $\epsilon$-orbit lies in the (001) plane through the center of the Brillouin zone (plotted in XCrySDen \cite{Kokalj1999:XCrySDen}).}
\label{fig:Fermisurface}
\end{figure}

We systematically compare Compton scattering and 2D-ACAR by performing measurements on the same disc-shaped Pd single crystal ($\varnothing = \SI{10}{\milli\meter}$, \SI{1}{\milli\meter} thick, surface normal [011]), serving as a model system with a well characterized electronic structure \cite{Brown1972:Magnetoacoustic, Mueller1970:Electronic}. During the 1970s, several theoretical and experimental studies focused on the electronic structure of Pd to determine the main features of the Fermi surface. Even today, it continues to attract the interest of scientists to explain subtler details such as the impact of electron-electron interactions \cite{Hayashi2013:High} or the exact shape of the L-hole Fermi surface pocket \cite{Oestlin2016:Electronic} (see figure \ref{fig:Fermisurface}). In this paper, we will not discuss specific features of the electronic structure of the system in detail. Here, 1D and 2D projections of the EMD and TPMD are compared in $\bm{p}$-space as well as in the reduced zone scheme of the first Brillouin zone ($\bm{k}$-space). Additionally, the experimental data are compared to first-principles calculations.

\section{Methods}
\subsection{Compton scattering}
In a Compton scattering experiment, a so-called 1D Compton profile, $J\left(p_z\right)$, is measured,
\begin{equation}
	J\left(p_z\right)=\iint\rho\left(\bm{p}\right)\mathrm{d}p_x\mathrm{d}p_y
\end{equation}
which is the twice-integrated 3D-EMD, $\rho\left(\bm{p}\right)$, where $p_z$ is the electron momentum along the scattering vector. From a quantum-mechanical point of view, $\rho\left(\bm{p}\right)$ would have to be expressed in terms of the excitations of the many-body wave function of the whole system. However, if treated in the independent-electron model and explicitly neglecting correlations, common in the Kohn-Sham formalism of density functional theory (DFT), $\rho\left(\bm{p}\right)$ can be written in terms of the momentum space wave functions, $\Phi_{\bm{k},j}\left(\bm{p}\right)$ \cite{Barbiellini2001:Treatment}, which are the Fourier transform of the real space wave functions, $\psi_{\bm{k},j}\left(\bm{r}\right)$, 
\begin{equation}
\label{EMD}
\rho\left(\bm{p}\right)=\sum_{\bm{k},j}n_{\bm{k},j}\bigl|\Phi_{\bm{k},j}\left(\bm{p}\right)\bigl|^2=\sum_{\bm{k},j}n_{\bm{k},j}\biggl| \int \psi_{\bm{k},j}\left(\bm{r}\right)\text{e}^{-i\bm{p}\cdot\bm{r}}\mathrm{d}\bm{r} \biggl|^2,
\end{equation}
where $n_{\bm{k},j}$ is the occupation of the electron in band $j$ with wave vector $\bm{k}$, and with $n_{\bm{k},j}=0$ and $n_{\bm{k},j}=1$ for completely unoccupied and fully occupied states, respectively. $\rho\left(\bm{p}\right)$ exhibits the crystal point symmetries, but does not have the (discrete) translational symmetry of the reciprocal lattice. To recover the translational symmetry, the Lock-Crisp-West (LCW) theorem \cite{Lock1973:Positron} can be applied to transform $\rho\left(\bm{p}\right)$ into the crystal momentum $\bm{k}$-space.

In this study, ten Compton profiles were measured along crystallographic directions spaced equally between $\Gamma$-X and $\Gamma$-K. The measurements were performed at room temperature on the Cauchois-type high-resolution Compton spectrometer at beamline BL08W, SPring-8, using synchrotron radiation with an incident energy of \SI{115}{\kilo\electronvolt} and a scattering angle of \SI{165}{\degree} \cite{Sakurai1998:High,Hiraoka2001:new,Sakurai2004:Cauchois}. The resolution of the spectrometer was determined to be \SI{0.14}{\au} (FWHM). For the measurement of the Compton profiles, the crystal was mounted on a rotating stage with the $[001]$ crystallographic direction oriented along the vertical rotation axis. For every profile, data was collected for at least \SI{380}{\minute{.}} leading to more than \num{1.6e5} counts in the Compton peak. Additionally, calibration measurements using a BiTl sample were made. Background spectra were taken along the $\Gamma$-X and $\Gamma$-K directions and halfway between them. The background for all other directions was linearly interpolated.

The data were corrected for several systematic effects to retrieve the Compton profiles from the measured data, in consecutive order: Saturation effects of the detector, pincushion-type geometrical correction, background, detection efficiency, multiple scattering \cite{Brancewicz2016:Monte}, and absorption inside the sample. Further details on the different corrections can be found elsewhere \cite{Brancewicz2013:Compton, Cooper2004:X}. Finally, the 1D-profiles are normalized to the number of electrons in the corresponding momentum range (determined from Hartree-Fock free-atom Compton profiles \cite{Biggs1975:Hartree}) and have units of electrons per atomic unit of momentum space (el./a.u.).

\subsection{Direct Inversion Method (DIM) in Compton scattering}
The reconstruction of the 2D projection of the EMD from 1D profiles is one way to compare Compton and 2D-ACAR experiments. The starting point for our approach is a linear operator, $\bm{T}^\alpha$, which projects the 2D plane projection, $\rho^{2D}$, of the EMD onto a 1D line spectrum, $y^\alpha$ (i.\,e. the Compton profiles), and thus mimics the Compton scattering experiment. One fact that we implicitly take into account is that a 3D-to-1D projection can be seen as successive projections from 3D to 2D and then from 2D to 1D. $\bm{T}^\alpha$ distributes the density in every pixel of $\rho^{2D}$ into the bins of $y^\alpha$ depending on the projection angle $\alpha$,
\begin{equation}
	y^\alpha=\bm{T}^\alpha \rho^{2D}.
\end{equation}
While $y^\alpha$ can be computed easily from a given $\rho^{2D}$, solving the inverse problem, i.\,e. finding $\rho^{2D}$ from a small number of projections, is more complex since it is under-determined. 

Many elements of $\rho^{2D}$  are equivalent due to the crystal point symmetry. Therefore, we can introduce the symmetry operator, $\bm{S}$, which reduces the dimension of the problem to the independent degrees of freedom, $x$, representing the irreducible area of the reconstructed plane projection,
\begin{equation}
	\rho^{2D}=\bm{S}x.
\end{equation}
By concatenating the $y^\alpha$ vectors and $T^\alpha$ matrices to a general measurement vector, $y$, and projection matrix, $\bm{T}$, respectively, the $\alpha$ index and summations can be neglected in the following. 
The maximum-likelihood estimation of the EMD $x$ can now be obtained by minimizing the $\chi^2$-functional,
\begin{equation}
	\chi^2\left(x\right)=\left(y-\bm{T}\bm{S}x\right)^\top\bm{W}\left(y-\bm{T}\bm{S}x\right).
\end{equation}
The weighting matrix, $\bm{W}$, is a diagonal matrix with the values $\sigma_i^{-2}$, where $\sigma_i$ are the uncertainties of each measured data point. 

Since the inverse problem is under-determined, a large number of different reconstructions $x$ may give $\chi^2$ values in accordance with the correct statistical accuracy. Additionally, were $\chi^2\left(x\right)$ to be minimized without any additional regularization, it is expected that the Poisson noise of the data would lead to noise in the reconstruction, commonly known as over-fitting. A regularization functional, $r\left(x\right)$, is therefore introduced to find a smooth and, thus, physically meaningful solution. This leads to the new functional,
\begin{equation}
	f\left(x\right)=\chi^2\left(x\right)+ r\left(x\right),
\end{equation}
which has to be minimized.

In Compton scattering, the momentum density decays only slowly with increasing momentum as all electron states (including the strongly localized core states) contribute to the measurement. Conventionally, a large momentum range thus has to be reconstructed in order to avoid reconstruction artefacts, implying either an excessive increase of the required computational power or a coarse sampling of momentum space. We solved this problem by utilizing the fact that the core levels, which are exclusively responsible for the densities at large momenta, have a signature that is isotropic for all practical purposes. This leads us to the following procedure: All spectra, $y^\alpha$, are averaged by a sum of Gaussians centered at zero momentum. This average is then subtracted from every Compton profile, giving the 1D anisotropies. From those, the 2D anisotropy is reconstructed and since the transformation of a Gaussian from 1D to 2D is known, the isotropic 2D-Gaussian functions can be added to the reconstruction of the anisotropy afterwards, which gives the reconstruction of the full signal.

As the anisotropy of the data comprises positive and negative values, the classical maximum entropy regularization \cite{Gull1978:Image} can not be applied in this case as the logarithm can only deal with positive densities. Additionally, as mentioned above, the regularization should penalize noise. However, while the 3D-EMD shows sharp steps at the Fermi surface, the 2D projection of the EMD is smeared. The projection integral in general leads to continuous 2D EMDs. To bias the reconstruction towards a corresponding behaviour, we use a sum of first and second derivatives as a regularization functional seems reasonable. The first derivative operator $\bm{D_1}$ can be expressed as a discrete difference operator, which calculates the difference between every pixel and the pixel to the right of it, and the difference between the pixel and the pixel below. Thus, if $\rho^{2D}$ has the dimension $m \times m$, $\bm{D_1}$ has the dimension $2m\left(m-1\right) \times m^2$. The second derivative operator $\bm{D_2}$ can easily be derived by multiplying $\bm{D_1}$ with its transpose matrix. To summarize, $f\left(x\right)$ can be written as follows,
\begin{equation}
\begin{split}
	 f\left(x\right)=&\chi^2\left(x\right)+ \lambda_1 x^\top\bm{S}^\top\bm{W_1}^\top\bm{D_1}^\top\bm{D_1}\bm{W_1}\bm{S}x \\
	  &+ \lambda_2 x^\top\bm{S}^\top\bm{W_2}^\top\bm{D_2}^\top\bm{D_2}\bm{W_2}\bm{S}x,
	 \end{split}
\end{equation}
where $\lambda_{1,2}$ are positive real numbers. As we expect higher variation of the anisotropy in areas with higher intensity, it makes sense to weigh the regularization functionals in a way that the relative variations are respected. Therefore, we use the 2D reconstruction of the isotropic part of the Compton profiles to calculate the weighting matrices $\bm{W_1}$ and $\bm{W_2}$, which are both the same, with the inverse of the isotropic 2D reconstruction along the diagonal.

To find the minimum of the quadratic function $ f\left(x\right)$ we have to find the $x$ which makes $\nabla f\left(x\right)=0$. The matrices associated to both the $\chi^2$ and the regularization functionals can be written as the square of real matrices (or a sum thereof) and are consequently positive semi-definite. The null space of the regularization functional matrix consists of the space of constant densities, which is not in the null space of the $\chi^2$ functional matrix, thus the quadratic form $f\left(x\right)$ is positive definite, its associated matrix is invertible, and the linear system of equations 
\begin{equation}
\begin{split}
	 0=&-2\bm{S}^\top\bm{T}^\top\bm{W}^\top\left(y-\bm{T}\bm{S}x\right)\\
	 &+2\lambda_1 \bm{S}^\top\bm{W_1}^\top\bm{D_1}^\top\bm{D_1}\bm{W_1}\bm{S}x\\ 
	 &+ 2\lambda_2 \bm{S}^\top\bm{W_1}^\top\bm{D_2}^\top\bm{D_2}\bm{W_1}\bm{S}x,
	 \end{split}
\end{equation}
has exactly one solution. This unique $x$ can easily be calculated by standard methods for solving linear systems of equations and is formally given by 
\begin{equation}
\begin{split}
	& x=\Bigg[ \bm{S}^\top\left(\bm{T}^\top\bm{W}\bm{T}+\lambda_1\bm{W_1}^\top \left(\bm{D_1}^\top\bm{D_1}\right)\bm{W_1} + \right.\Bigg. \\
	& \Bigg.\left.\lambda_2\bm{W_1}^\top\left(\bm{D_1}^\top\bm{D_1}\bm{D_1}^\top\bm{D_1}\right)\bm{W_1}\right) \bm{S}\Bigg]^{-1} \left(\bm{S}^\top\bm{T}^\top\bm{W}y\right).
\end{split}
\end{equation}
This makes the DIM reconstruction technique fast and efficient in comparison to reconstruction algorithms applying non-linear regularization functionals. Furthermore, in case of the iterative methods, a convergence criterion has to be defined. This is not necessary for the DIM as the inversion directly gives the final result.

\subsection{2D-ACAR}
2D-ACAR exploits the annihilation of electron-positron pairs to investigate the electronic structure \cite{Weber2015:Spin, Weber2017:Electronic, West1995:Positron, Ceeh2016:Local}. After implantation into the sample, positrons thermalize within a few picoseconds and subsequently propagate in Bloch states, as they also feel the crystal potential. The annihilation of these Bloch-state positrons with electrons leads predominately to the emission of two photons. In the center-of-mass (rest) frame of the electron-positron pair, the photons are emitted in opposite directions with equal energy due to conservation of momentum. However, in the laboratory frame, the transverse (i.e. perpendicular to the emission direction of the photons) components of the electron and positron momenta lead to a proportional deviation of the emission angle from \SI{180}{\degree} (the small angle approximation is valid because the electron and positron momenta are small in comparison with the momentum of a \SI{511}{keV} photon).

\begin{figure*}
\includegraphics[width=\textwidth]{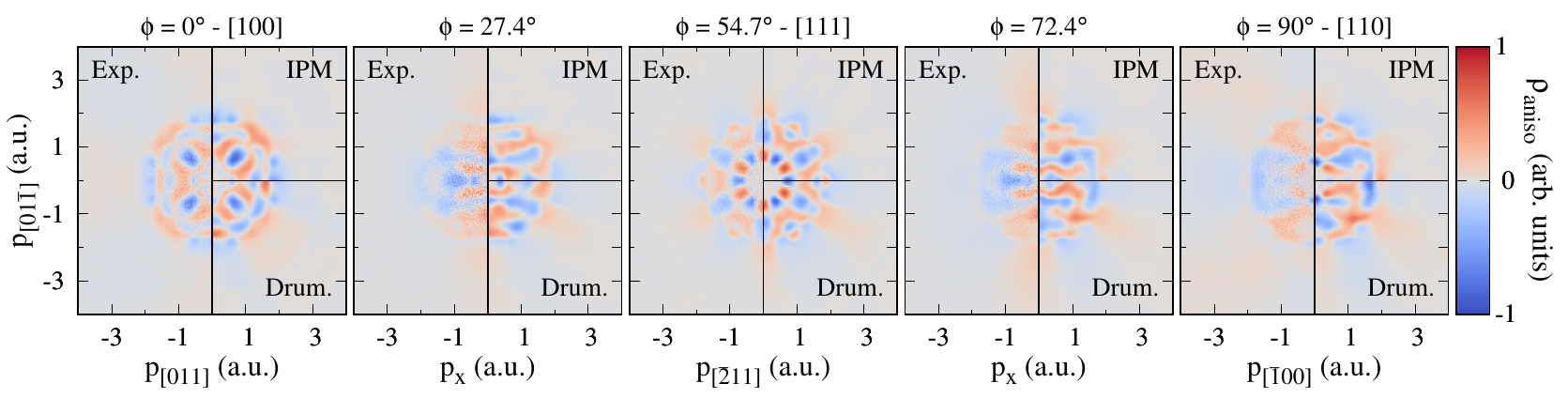}
\caption{2D radial anisotropy of the five measured ACAR spectra (left half of each subplot) and the corresponding theoretical calculation, with (bottom right) and without Drummond (top right) enhancement. All calculations have been convolved with the experimental resolution, and all experimental spectra were symmetrized to the according crystal symmetry. The angle $\Phi$ denotes the angle measured from the [100] direction within the (011) plane.}
\label{fig:aniso_2D-ACAR_all_directions}
\end{figure*}

The coincident measurement of the angular distribution of the two gamma quanta yields the projection of  the two-photon momentum density (TPMD), $\rho^{2\gamma}$. Neglecting electronic correlations, the TPMD can be expressed as follows,
\begin{equation}
\label{TPMD}
\rho^{2\gamma}\left(\bm{p}\right)=\sum_{\bm{k},j}n_{\bm{k},j}\biggl| \int \sqrt{\gamma\left(\bm{r}\right)}\psi_{\bm{k},j}\left(\bm{r}\right)\psi_+\left(\bm{r}\right)\text{e}^{-i\bm{p}\cdot\bm{r}}\mathrm{d}\bm{r} \biggl|^2,
\end{equation}
which is closely related to the definition of  the EMD in equation \ref{EMD}. They only differ by the positron wave function, $\psi_+\left(\bm{r}\right)$, and electron-positron correlations described by the so-called enhancement factor \cite{Sterne1991:First, Drummond2011:Quantum}, $\gamma\left(\bm{r}\right)$. The TPMD can be thought of as the EMD {\it as seen by the positron}. When no electron-positron enhancement is included ($\gamma=1$), the calculation is described as being in the independent-particle model (IPM).

The positron experiments were performed on the 2D-ACAR spectrometer at Technische Universität München, which offers a detector-detector distance of \SI{17.5}{\meter}. For further details on the experimental setup, we refer to Ref. \cite{Ceeh2013:source}. In total, five measurements within the $(011)$ plane were performed. The projections were taken along the crystallographic cubic high symmetry directions $[100]$, $[0\bar{1}1]$ and $[1\bar{1}1]$, and at \SI{27.4}{\degree} and \SI{72.4}{\degree} away from $[100]$ in the $(011)$ plane. All measurements were made at \SI{10}{\kelvin} to minimize the resolution degradation due to thermal motion of the positron, and more than \num{5e7} coincident counts were collected per spectrum. Owing to the relatively short detector-detector distance, the experimental resolution (FWHM) was \num{0.21} and \SI{0.17}{\au} in the $p_{\rm horizontal}$ and $p_{\rm vertical}$ directions, respectively \cite{Leitner2012:Eliminating}. Both momentum directions are perpendicular to the detector-detector direction.

\subsection{First-principles electronic structure calculation}

In order to compare with the experimental momentum densities, ground-state electronic structure calculations were performed.
The \textsc{Elk} code \cite{Elk:Link}, a highly-accurate full-potential augmented plane-wave plus local orbital (FP-APW+lo) method, was used to calculate the ground-state electronic structure of face-centered-cubic Pd at the experimentally-determined \cite{Miiller1971:Crystal} cubic lattice constant, $a=\SI{3.890}{\angstrom}$.
Convergence was achieved with a $16\times16\times16$ ${\bm k}$-point grid with a plane-wave cut-off in the interstitial region of $|{\bm G}+{\bm k}|_{\rm max}=8.0/R_{\rm mt}$ (where $R_{\rm mt}=\SI{2.57}{\au}$ was the muffin-tin radius) and the Perdew-Burke-Ernzerhoff generalised gradient approximation (PBE-GGA) \cite{Perdew1996:Generalized} was used for the exchange-correlation functional. The valence electron configuration was $4s^2 4p^6 4d^{10}$, and the remaining 28 electrons were considered to be core.
Since Pd has a relatively high atomic number, the spin-orbit interaction was included in the calculations by adding a term of the form $\boldsymbol\sigma\cdot{\bm L}$ (where $\boldsymbol\sigma$ is the spin vector and ${\bm L}$ is the orbital angular momentum vector) to the second variational Hamiltonian.
Because the Compton scattering and 2D-ACAR experiments were performed at room temperature and at $T=\SI{10}{\kelvin}$, respectively, smearing widths (effective electronic temperatures) of \SI{300}{\kelvin} and \SI{30}{\kelvin} were used in the ground-state calculations from which the EMD and TPMD, respectively, were calculated using the method of Ernsting \textit{et al.} \cite{Ernsting2014:Calculating}.
Compton scattering is equally sensitive to \textit{all} electrons (core and valence) and, while the EMD was calculated only for the valence electrons, the momentum cut-off was $|{\bm p}|_{\rm max}=16.0$~a.u. to include contributions from the most tightly bound semi-core valence states. In the case of the TPMD this cut-off could be reduced to $|{\bm p}|_{\rm max}=8.0$~a.u. due to the small overlap of the positron wave function with the more tightly bound electron states.
In order to understand the effect of electron-positron correlations on the measured densities, two different TPMD calculations were performed with the IPM, namely one assuming no enhancement [$\gamma \left(\bm{r}\right)=1$] and a second applying the positron enhancement model proposed by Drummond \textit{et al.} \cite{Drummond2011:Quantum} with gradient corrections \cite{Barbiellini1995:PositronGGA}.

\section{Results and Discussion}

\subsection[2D-ACAR and 1D Compton scattering: Experiment and Theory]{2D-ACAR and 1D Compton scattering: \\ Experiment and Theory}

First, we compare the experimental results with the corresponding theoretical calculations, namely the experimental 2D-ACAR data to TPMD calculations and Compton data to EMD calculations. One of the first useful quantities to consider is the radial anisotropy in which the isotropic average density (averaged on circles of fixed momentum) is subtracted from the density itself. In the case of a simple metal, this radial anisotropy can be dominated by the presence of the Fermi surface but it also contains information about the anistropy of the wavefunctions of electrons in filled bands. Figure \ref{fig:aniso_2D-ACAR_all_directions} shows the radial anisotropy of the five 2D-ACAR measurements and the corresponding theoretical TPMD calculations (IPM and Drummond enhancement). The theoretical spectra are convolved with a two-dimensional Gaussian function accounting for the instrumental momentum resolution. All experimental spectra were symmetrized according to the expected crystal symmetry. Comparing the calculation with Drummond \cite{Drummond2011:Quantum} enhancement to the IPM, we can see that the enhancement generates stronger anisotropies at higher momenta due to the fact that the positron, which is screened by an electron cloud, has an increased overlap with the more tightly bound electrons (which contribute at larger momentum) due to the weaker Coulomb repulsion \cite{Rubaszek2008:EPcorrelation}. Overall we can state that, while there are regions of the experimental data which agree more closely with either one or the other approximations, it is certainly not the case that the enhancement produces a significant improvement in the radial anisotropy.

\begin{figure}
\includegraphics[scale=1]{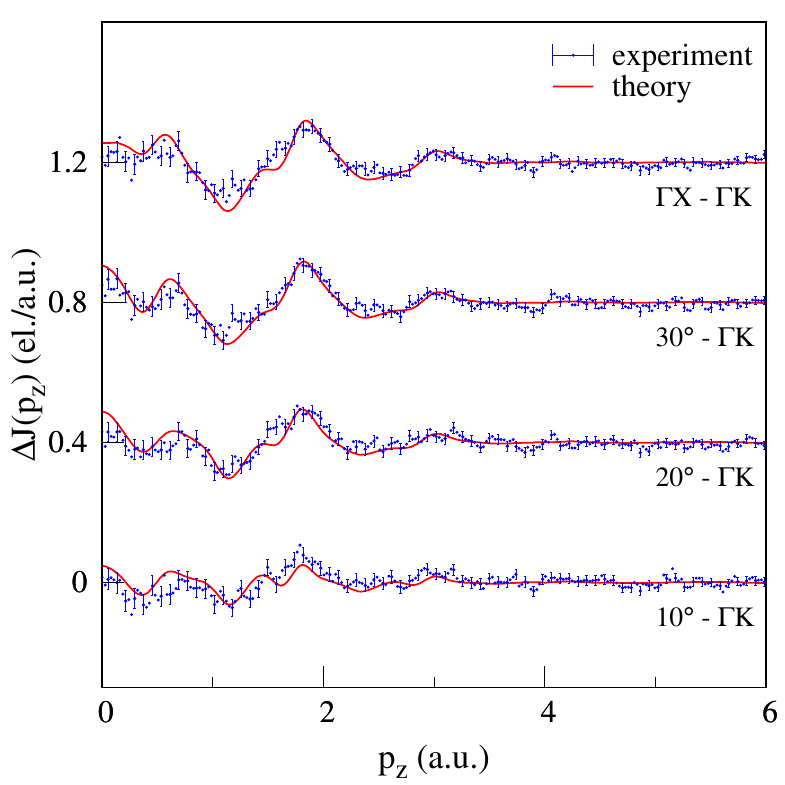}
\caption{Directional differences of experimental (blue) and theoretical Compton profiles (red). The labelled angles are measured from the $\Gamma - K$ direction towards the $\Gamma - X$ direction of the fcc Brillouin zone. The calculated profiles are convolved with a one-dimensional Gaussian accounting for the experimental resolution. For every third experimental data point an errorbar showing the statistical error of one standard deviation is plotted. The plots have been offset by \SI{0.4}{el{.}\per\au} from one another for clarity.}
\label{fig:Compton_DirDif}
\end{figure}

Figure \ref{fig:Compton_DirDif} shows the directional difference between four Compton profiles measured along different directions and the Compton profile measured along the $\Gamma$-K direction. In all but the lowest $Z$ elemental metals, the anisotropy between directions is usually dominated by the electrons in filled bands (since there are many more of them), rather than the (small number of) partially filled bands which give rise to the Fermi surface. All measurements show very good agreement with the first-principles calculations (which have been convolved with a one-dimensional Gaussian accounting for the experimental resolution). This is also true for the other five directions which are not explicitly shown here. The uncertainties are calculated from counting statistics propagated through the corrections.

\subsection{2D-Reconstruction from 1D-Compton Profiles}

\begin{figure}

\subfigure[]{\includegraphics[scale=1]{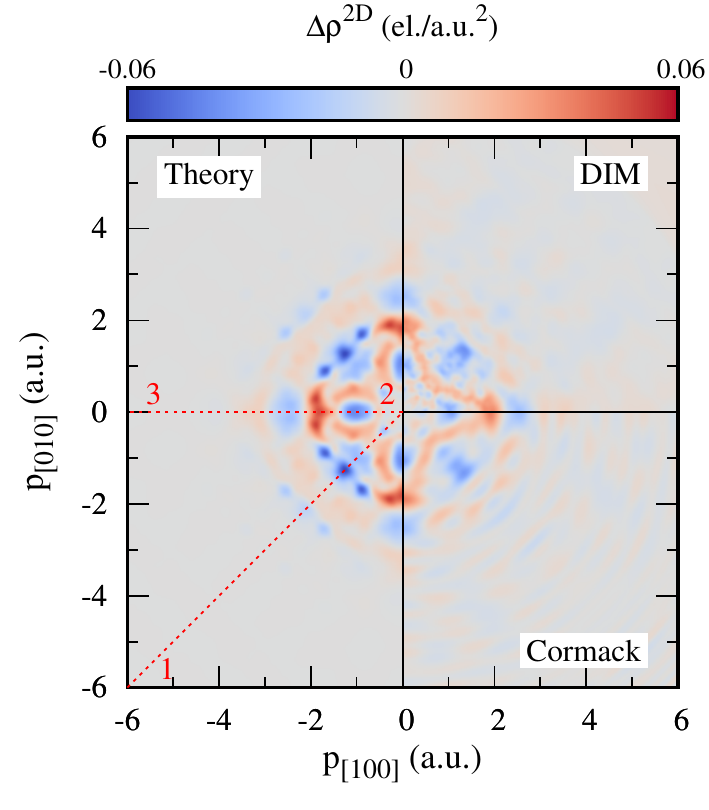}}

\subfigure[]{\includegraphics[scale=1]{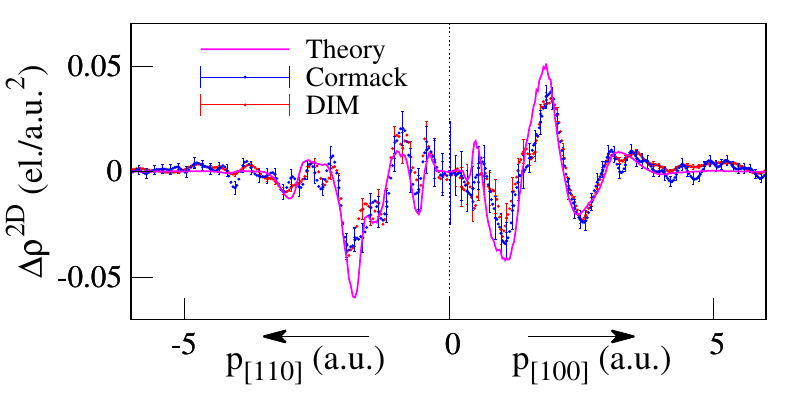}}
\caption{(a) 2D radial anisotropy of the projected EMD calculated by DFT (left), reconstructed from ten 1D Compton profiles by DIM (top right) and by Cormack (bottom right). The theoretical spectrum was convolved with a two-dimensional Gaussian accounting for the experimental resolution. (b) Cut through the 2D radial anisotropy along the red path (1$\rightarrow$2$\rightarrow$3) shown in (a).}
\label{fig:anisoRek}
\end{figure}

As the Compton experiment measures 1D projections of the EMD, the 2D projection has to be reconstructed from a series of 1D measurements to compare it with the 2D-ACAR measurements. This was achieved with both the new DIM algorithm and the well-known Cormack method \cite{Kontrym-Sznajd1990:Three} in order to benchmark the new algorithm. The required computational effort is much higher in DIM compared to Cormack due to the high number of free parameters in the DIM. However, a standard PC is still capable of calculating a reconstruction of $512 \times 512$ pixels from the ten Compton profiles within several minutes. In order to compare the results of both algorithms, it is, again, useful to consider the radial anisotropy of the (projected) EMD. This anisotropy will have contributions from both filled (due the anisotropy of the electron wavefunctions) and from partially filled bands (in which case it additionally contains information about the Fermi surface). Figure \ref{fig:anisoRek}(a) shows the radial anisotropy of the reconstructed spectra and of the first-principles calculations. All of the main features of the theoretical spectrum are reconstructed comparably well by both methods. At high momenta, the noise of the DIM reconstruction is more isotropic and, compared to Cormack, exhibits smaller variation in the radial direction than the tangential direction.
Figure \ref{fig:anisoRek}(b) shows a cut through the 2D anisotropy along high-symmetry directions according to the red path (1$\rightarrow$2$\rightarrow$3) shown in (a). This cut highlights more clearly the good agreement of both reconstructions within the error bars.

\begin{figure}
\centering
\includegraphics[scale=1]{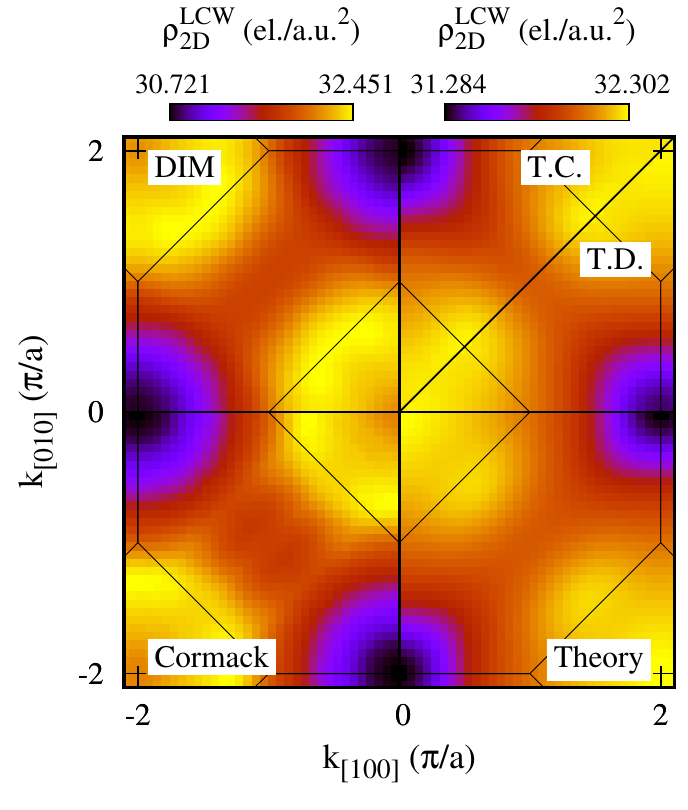}
\caption{2D LCW of the experimental and theoretical  2D-EMD: Reconstructed experimental spectrum by DIM (top left) and reconstructed experimental data by Cormack (bottom left); theoretical EMD (bottom right); reconstruction from theoretical 1D Compton profiles by Cormack (top right, T.C.) and DIM (top right, T.D.). The theoretical data in the right half of the plot was convolved with a two-dimensional Gaussian accounting for the experimental resolution, before back-folding.}
\label{fig:lcwRecon}
\end{figure}

The LCW back-folded data, presented in figure \ref{fig:lcwRecon}, is almost identical for both reconstruction methods. One interesting difference between theory (bottom right) and experiment (left) is the intensity distribution around the center (projected $\Gamma$-point), where the theory shows a high intensity while both reconstructions give a clear dip. The possibility of this behavior being an artefact of the reconstruction was excluded by reconstructing the 2D LCW from theoretically calculated 1D Compton profiles (top right), which did not show a dip in the center of the LCW. 
Typically, theoretical calculations using the LDA or GGA are not fully capable of reproducing all parts of the Fermi surface equally well \cite{Harrislee2021:Sensitivity}. From our Fermi surface calculations (figure \ref{fig:Fermisurface}), we numerically extracted multiple dHvA orbits using the SKEAF code \cite{Rourke2012:Numerical}. Most of the orbits agree well with dHvA measurements \cite{Dye1981:Fermi} and the area of the so-called $\epsilon$-orbit, which originates from a heavy electron band with an effective cyclotron mass of \SI{12.5}{m_e} agrees with our calculations within \SI{1}{\percent} \cite{Vuillemin1999:High}. However, the area of the so-called $\alpha$-orbit is about \SI{11.5}{\percent} smaller in DFT in comparison to the dHvA experiment \cite{Dye1981:Fermi}. As this orbit also originates from a heavy band, such differences might be expected as a slight change in the position of the band relative to the Fermi energy and can hence strongly influence the Fermi surface created by the band. To get a feeling of the size of the Fermi surface tube corresponding to the $\alpha$-orbit in our Compton experiment, we calculated the first derivatives of cuts through the LCW along the $\Gamma$-X and $\Gamma$-K directions. Both curves indicate a hole pocket at the projected X-points which is larger than expected from theory. Therefore, we attribute the dip in the experimental LCW at the center of the projected Brillouin zone (where the X-points also projects) to this difference between theory and experiment.

\subsection{Comparison of ACAR and Compton (in 2D and 1D)}
Now, we compare the results of ACAR and Compton using the new DIM algorithm. 

\begin{figure}
\centering
\includegraphics[scale=1]{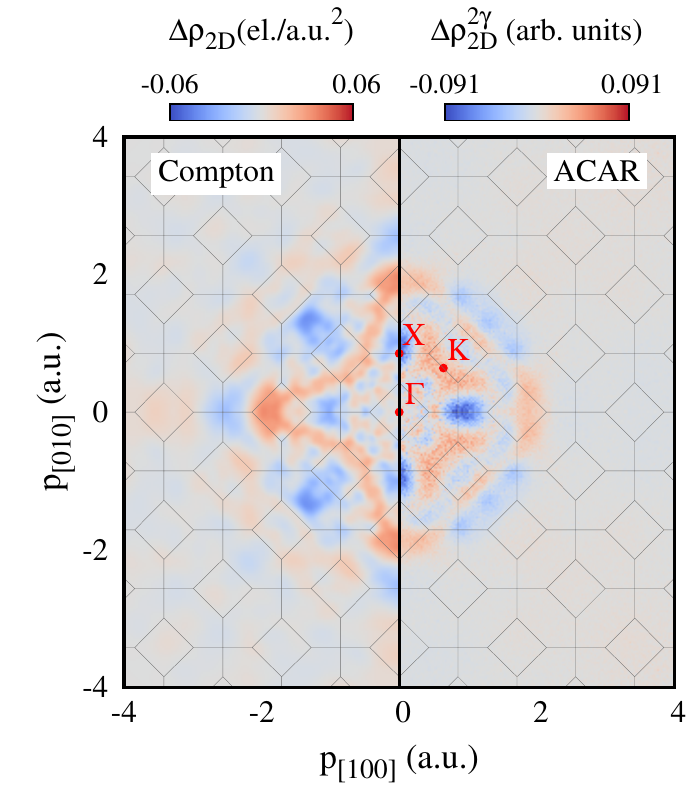}
\caption{2D radial anisotropy of the EMD reconstructed (DIM) from ten 1D Compton spectra (left) and the corresponding 2D projection obtained by 2D-ACAR (right). Both spectra were normalized to the corresponding electron density of the EMD before calculating the radial anisotropy. The positions of the projected high-symmetry $\Gamma$, K, and X points are indicated.}
\label{fig:anisoMeas}
\end{figure}

Figure \ref{fig:anisoMeas} shows the experimental radial anisotropy of the DIM-reconstructed 2D-EMD (from Compton) and the 2D-TPMD (from 2D-ACAR). The labelled $\Gamma$, K, and X points are the projected positions of the high symmetry points of the three-dimensional face-centered cubic Brillouin zone. The 2D-ACAR shows significantly larger anisotropy, especially at low momenta, compared to Compton scattering, which is expected
because the positron wavefunction overlaps strongly with that of the delocalized electrons (particularly with electrons at the Fermi surface), but only overlaps very weakly with that of the most tightly bound states. Most of the main features like the low intensity pocket around X and the butterfly shaped high intensity around the K-point are revealed by both techniques. However, obvious differences are also present, particularly at higher momenta, e.\,g. for $|{\bm p}|>\SI{2}{\au}$, where 2D-ACAR hardly reveals any anisotropy, again due to the wavefunction overlap.

\begin{figure}
\centering
\includegraphics[scale=1]{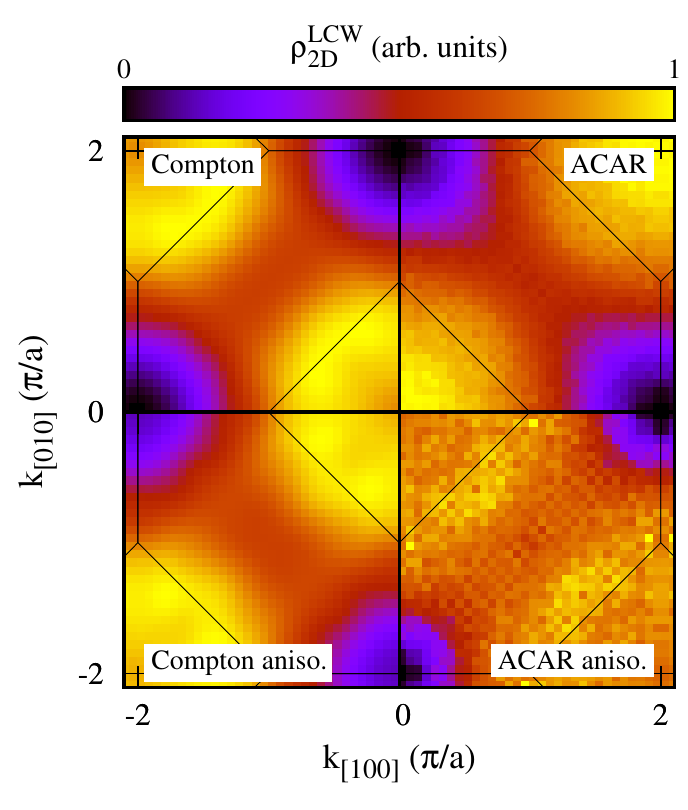}
\caption{2D-LCW of the reconstructed (DIM) Compton spectrum (top left) and the 2D-ACAR spectrum (top right). 2D-LCW calculated from the radial anisotropies of the reconstructed (DIM) Compton spectrum (bottom left) and the 2D-ACAR spectrum (bottom right).}
\label{fig:lcw2D}
\end{figure}

\begin{figure}
\centering
\includegraphics[scale=1]{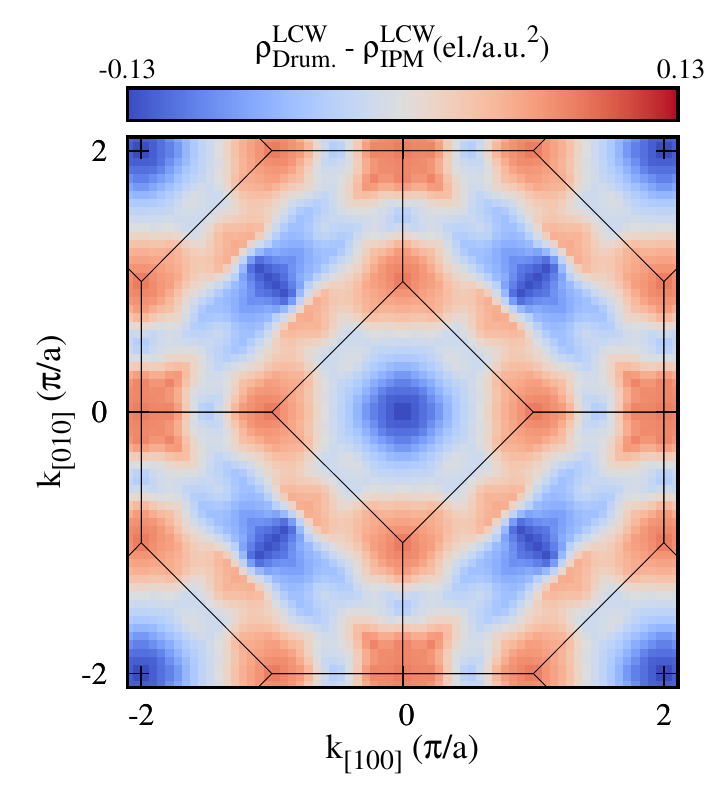}
\caption{Difference between the TPMD calculation including enhancement (Drummond model \cite{Drummond2011:Quantum}) and the IPM calculation in the reduced zone scheme.}
\label{fig:drumvsipm}
\end{figure}

For a further comparison, the spectra are back-folded into the first Brillouin zone \cite{Lock1973:Positron}. The Compton LCW is normalized, so that the integral over one Brillouin zone equals 18 valence electrons and a constant (${\bm k}$-independent) contribution of 28 core electrons is added, since Compton scattering probes every electron equally. Although this normalization is not technically valid for the TPMD measured by 2D-ACAR, the LCW is also normalized in the same way for easier comparison. In the upper half of Fig.~\ref{fig:lcw2D}, the LCWs of the experimental 2D-EMD and 2D-TPMD are shown on the left and right side, respectively.
In contrast to the Compton experiment, the 2D-ACAR experiment does not show a dip in the center of the Brillouin zone. This is not expected from positron enhancement effects because the difference between the theoretical calculation including positron enhancement and the calculation using the IPM suggests a reduced intensity at the projected $\Gamma$-point (see figure \ref{fig:drumvsipm}). We attribute the increased LCW back-folded density at low momenta to a contribution of positrons annihilating in vacancy-type defects. As shown by Dugdale and Laverock \cite{Dugdale2014:Recovering}, the Fermi surface information can still be recovered by instead back-folding the radial anisotropy to the first Brillouin zone instead of calculating the LCW from the full density. This is shown for Compton data (left) and 2D-ACAR data (right) in the lower half of Fig.~\ref{fig:lcw2D}. As depicted, the LCW of the back-folded anisotropies agree very well over the whole Brillouin zone, including the zone center. 

\begin{figure}
\centering
\includegraphics[scale=1]{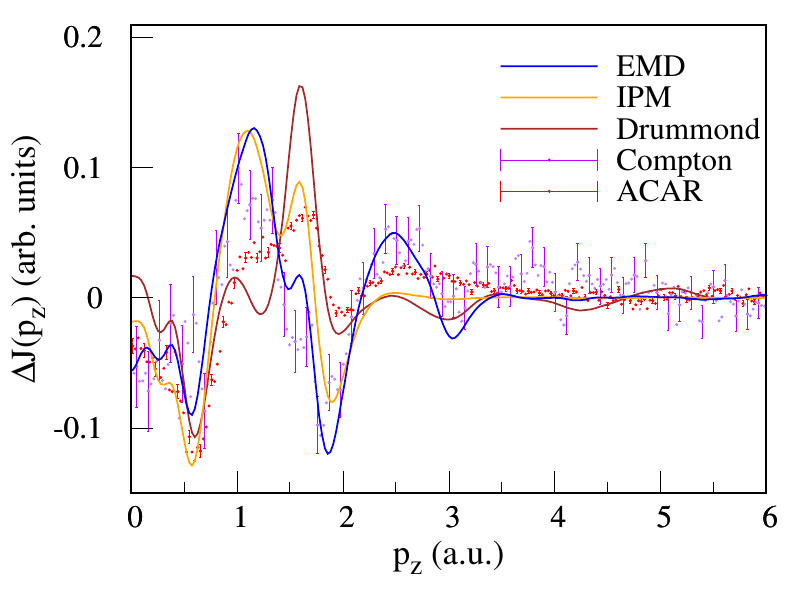}
\caption{Directional differences between 1D projections along $\Gamma$-X and $\Gamma$-K of the Compton and 2D-ACAR experiments (purple and red points, respectively), together with the calculated (lines) EMD (blue) and TPMD, calculated with (brown) and without (orange) positron enhancement. The calculated profiles are convolved with Gaussian functions accounting for their respective experimental resolutions. For every fourth experimental data point, error bars show the statistical error of one standard deviation.}
\label{fig:Dif_Elk}
\end{figure}

To get a more quantitative comparison between 2D-ACAR and Compton scattering, 1D projections of the the 2D-ACAR measurements along the $\Gamma$-K and $\Gamma$-X directions are retrieved by summation of the $[011]$ projection along the horizontal and vertical measurement directions, respectively. In order to visualize details of anisotropic features of the distributions more clearly, the directional differences of the two 1D profiles are calculated. The results are shown in figure \ref{fig:Dif_Elk} for Compton and 2D-ACAR measurements as well as for the EMD and TPMD (IPM and including enhancement) calculations. First, comparing the results of the DFT calculations, as expected, we can see significant differences between EMD and TPMD, which can be attributed to the influence of the positron. The same holds true if we compare the ACAR and Compton measurements, especially in the region from \SIrange{1}{2.2}{\au}. Furthermore, we can clearly see an enhancement effect on the TPMD calculations by comparing the results from the IPM and the calculation using the Drummond enhancement model. Comparing experimental data to theory we see that the Compton experiment is very well described by the calculated EMD. The 2D-ACAR data is not equally well described by either of the two TPMD models over the whole momentum range. At low momenta and around \SI{1.6}{\au} the data is better described by the IPM, while, at other momenta, the Drummond model seems to deliver the better approximation.
This clearly shows how strongly TPMD calculations and 2D-ACAR experiments are influenced by positron wave function and enhancement effects that make theoretical modelling of positron spectra much more difficult. 

\section{Conclusion}
We performed Compton scattering and 2D-ACAR measurements on a high-quality Pd single crystal in order to compare the results from both experiments and reveal the influence of positron probing effects on the measured electronic structure. In order to allow a reliable comparison of the 2D projections of the TPMD from 2D-ACAR with the 1D Compton profiles, a new, direct reconstruction technique was developed for the 1D-to-2D reconstruction of Compton data. Our DIM algorithm uses the direct inversion of linear matrices and is a general case of Thikonov regularization to solve the reconstruction problem. The results from the DIM agree well with a reconstruction by the well-known Cormack method. Even if the new approach is computationally more demanding than the Cormack's method, with modest computational power, the DIM algorithm still enables an efficient method to get a high quality reconstruction of the 2D electron momentum density. 

In order to tackle the 3D reconstruction problem from either 1D Compton data or 2D-ACAR data the application of the DIM seems reasonable as it is based on a very universal approach. It could offer some advantages like the easy inclusion of the experimental resolution function into the reconstruction algorithm or more freedom in choosing your projection directions during the experiment. However, some caution has to be taken in choosing an appropriate regularization functional. Since the first derivative regularization would probably lead to a relatively smooth 3D density, as it does in other approaches like the Hermite polynomials or spherical harmonics, the utilisation of the zeroth order derivative might be a good choice.

Differences between theory and experiment in the LCW back-folded spectra support earlier findings by dHvA experiments that the DFT calculations underestimate the size of the of the so called $\alpha$-orbit.

In order to analyze the influence of positron probing effects on the determination of the electronic structure, first-principles calculations of the EMD and TMPD were performed. For the TPMD, clear differences between both models (namely, the IPM and the Drummond enhancement model) can be found, however, neither are fully capable of describing the experimental data over the whole momentum range.

A huge advantage of 2D-ACAR is the direct measurement of a 2D projection of the TMPD compared to a 1D projection of the EMD measured in Compton scattering. This drawback can be compensated by reconstruction of the 2D information from 1D Compton profiles along different directions. Although in this work an efficient reconstruction technique was used, the data treatment including the reconstruction of the 2D spectrum needed in Compton scattering is much more demanding compared to 2D-ACAR. Besides the fact that Compton scattering is practically insensitive to vacancy-type defects, the biggest advantage of Compton scattering is the much simpler calculation of theoretical spectra, compared to the calculation of 2D-ACAR spectra in which enhancement and positron wave function effects, which are  difficult to calculate, might play an important role. This can be clearly seen in the directional differences between the EMD and TPMD calculations, as well as the convincing agreement between EMD and Compton measurements, while the positron experiment and theory show obvious discrepancies.

\begin{acknowledgments}
We are grateful to the authorities of SPring-8 (JASRI), Japan for granting beam time under the proposal No. 2016B1685. This project is funded by the Deutsche Forschungsgemeinschaft (DFG) within the Transregional Collaborative Research Center TRR80 “From electronic correlations to functionality”.
We acknowledge the support of the Supercomputing Wales project, which is part-funded by the European Regional Development Fund (ERDF) via Welsh Government.
\end{acknowledgments}

\bibliography{literature.bib}

\end{document}